\documentclass[usenatbib,letters]{mn2e}

\usepackage{natbib}
\usepackage{longtable,lscape}
\usepackage{amsmath}
\bibpunct{(}{)}{;}{a}{}{,}
\usepackage{graphicx}

\usepackage{graphicx}
\usepackage[colorlinks=true,citecolor=blue]{hyperref}
\usepackage{color}
\usepackage{xspace}
\usepackage{dcolumn}
\usepackage{longtable}
\usepackage{lscape}
\usepackage{url}
\usepackage{afterpage}

\newcommand{\MJup}{M$_{\mathrm{Jup}}$\xspace}
\newcommand{\RJup}{R$_{\mathrm{Jup}}$\xspace}

\newcommand{\MEarth}{M$_{\oplus}$\xspace}

\newcommand{\mic}{$\mu$m\xspace}
\newcommand{\as}{\hbox{$^{\prime\prime}$}\xspace}

\title[]{Upper limits for Mass and Radius of objects around Proxima Cen from SPHERE/VLT}
\author[D. Mesa et al.]{D. Mesa$^{1}$
  , A. Zurlo$^{2,3,4}$, J. Milli$^{7}$, R. Gratton$^{1}$, S. Desidera$^{1}$, M. Langlois$^{8,4}$, \newauthor
  A. Vigan$^{4}$, M. Bonavita$^{9,1}$, J. Antichi$^{1,22}$, H. Avenhaus$^{3,11,12}$, A. Baruffolo$^{1}$, B. Biller$^{9,12}$, \newauthor
  A. Boccaletti$^{13}$, P. Bruno$^{14}$, E. Cascone$^{15}$, G. Chauvin$^{5,6}$, R.U. Claudi$^{1}$, V. De Caprio$^{15}$, \newauthor
  D. Fantinel$^{1}$, G. Farisato$^{1}$, J. Girard$^{7}$, E. Giro$^{1}$, J. Hagelberg$^{5,6}$, S. Incorvaia$^{17}$, \newauthor
  M. Janson$^{18}$, Q. Kral$^{19}$, E. Lagadec$^{20}$, A.-M. Lagrange$^{5,6}$, L. Lessio$^{1}$, M. Meyer$^{11}$, \newauthor
  S. Peretti$^{16}$, C. Perrot$^{13}$, B. Salasnich$^{1}$, J. Schlieder$^{21,12}$, H.-M. Schmid$^{11}$, S. Scuderi$^{14}$, \newauthor
  E. Sissa$^{1,23}$, C. Thalmann$^{11}$, M. Turatto$^{1}$\\
$^{1}$INAF-Osservatorio Astronomico di Padova, Vicolo dell'Osservatorio 5, Padova, Italy, 35122-I\\
  $^{2}$N\'{u}cleo de Astronom´ıa, Facultad de Ingenier\'{i}a, Universidad Diego Portales, Av. Ejercito 441, Santiago, Chile\\
  $^{3}$Universidad de Chile, Camino el Observatorio, 1515 Santiago, Chile\\
  $^{4}$Aix Marseille Universit´e, CNRS, LAM - Laboratoire d’Astrophysique de Marseille, UMR 7326, 13388, Marseille, France\\
  $^{5}$Universit\'{e} Grenoble Alpes, IPAG, 38000 Grenoble, France\\
  $^{6}$CNRS, IPAG, 38000 Grenoble, France\\
  $^{7}$European Southern Observatory (ESO), Alonso de C\'{o}rdova 3107, Vitacura, 19001 Casilla, Santiago, Chile\\
  $^{8}$CRAL, UMR 5574, CNRS, Universit\'{e} Lyon 1, 9 avenue Charles Andr\'{e}, 69561 Saint Genis Laval Cedex, France\\
  $^{9}$Institute for Astronomy, University of Edinburgh, Blackford Hill View, Edinburgh EH9 3HJ, UK\\
  $^{10}$Millenium Nucleus Protoplanetary Disks in ALMA Early Science, Universidad de Chile, Casilla 36-D, Santiago, Chile\\
  $^{11}$ETH Zurich, Institute for Astronomy, Wolfgang-Pauli-Strasse 27, 8093 Zurich, Switzerland\\
  $^{12}$Max-Planck-Institut für Astronomie, K\"{o}nigstuhl 17, D-69117 Heidelberg, Germany\\
  $^{13}$LESIA, Observatoire de Paris-Meudon, CNRS, Universit\'{e} Pierre et Marie Curie, \\ Universit\'{e} Paris Diderot, 5 Place Jules Janssen, F-92195 Meudon, France\\
  $^{14}$INAF-Osservatorio Astrofisico di Catania, Via S. Sofia 78, I-95123 Catania, Italy\\
  $^{15}$INAF – Osservatorio Astronomico di Capodimonte, Via Moiariello,16 I-80131 Napoli, Italy\\
  $^{16}$Observatoire de Gen\'{e}ve, University of Geneva, 51 Chemin des Maillettes, 1290, Versoix, Switzerland\\
  $^{17}$INAF-Istituto di Astrofisica Spaziale e Fisica Cosmica di Milano, Via E. Bassini 15, 20133 Milano, Italy\\
  $^{18}$Department of Astronomy, Stockholm University, SE-106 91 Stockholm, Sweden\\
  $^{19}$Institute of Astronomy, University of Cambridge, Madingley Road, Cambridge CB3 0HA, UK\\
  $^{20}$Laboratoire Lagrange (UMR 7293), UNSA, CNRS, Observatoire de  la  C\^{o}te  d'Azur, Bd. de l'Observatoire, 06304 Nice Cedex 4, France\\
  $^{21}$NASA Exoplanet Science Institute, California Institute of Technology, Pasadena, CA, USA\\
  $^{22}$INAF-Osservatorio Astrofisico di Arcetri  L.go E. Fermi 5, 50125 Firenze, Italy\\
  $^{23}$Dipartimento di Fisica e Astronomia "G. Galilei", Universit\'{a} degli Studi di Padova, Vicolo dell’Osservatorio 3, 35122 Padova, Italy
}
\begin{document}
\date{Accepted . Received ; in original form }
\pagerange{\pageref{firstpage}--\pageref{lastpage}} \pubyear{}
\maketitle
\label{firstpage}
\begin{abstract}
  The recent discovery of an earth-like planet around Proxima Centauri has drawn much attention to this star and its environment. We performed a series of observations
  of Proxima Centauri using SPHERE, the planet finder instrument installed at the ESO Very Large Telescope UT3, using its near infrared modules, IRDIS and IFS.
  No planet was directly detected but we set upper limits on the mass up to 7\,au exploiting the AMES-COND models. Our IFS observations reveal that no planet more massive
  than  $\sim$ 6-7\,\MJup can be present within 1\,au. The dual band imaging camera IRDIS also enables us to probe larger separations than the other
  techniques like the radial velocity or astrometry. We obtained mass limits of the order of 4\,\MJup at separations of 2\,au or larger
  representing the most stringent mass limits at separations larger than 5\,au available at the moment. We also did an attempt to estimate the radius of possible
    planets around Proxima using the reflected light. 
  Since the residual noise
  for this observations are dominated by photon noise and thermal background, longer exposures in good observing conditions could further improve the achievable contrast limit.
\end{abstract}

\begin{keywords}
Instrumentation: spectrographs - Methods: data analysis - Techniques: imaging spectroscopy - Stars: planetary systems, Proxima Centauri
\end{keywords}

\section{Introduction}

After the recent discovery of a terrestrial planet around the star Proxima Centauri \citep{anglada} a new interest arose in the nearest star system to the Sun.
While this planet, that has a separation of just 0.05\,au with a period of 11.2 days and a minimum mass of 1.3\,\MEarth, cannot be imaged with the current
instrumentation aimed to detect the emitted light from extrasolar planets like e.g. GPI \citep{Mac06} and SPHERE \citep{2008SPIE.7014E..18B}, it would be however interesting to
have informations about further possible objects at larger separations to fully characterize the system. Exploiting direct imaging observations it is possible to put some
constraints on the mass and on the radius of other objects in the Proxima system.
A similar work has been done in the past exploiting both the radial velocity (RV) technique \citep{2008A&A...488.1149E,2009A&A...505..859Z,2014MNRAS.439.3094B}
and astrometric measurements \citep{2014AJ....148...91L}, but never exploiting direct imaging techniques. We repeatedly observed Proxima with SPHERE
in the past months with the aim to obtain precise astrometry of a background star which is undergoing a microlensing
event caused by the approaching of Proxima \citep{2014ApJ...782...89S}. This star is clearly visible even when is not undergoing the microlensing effect.
This will give a unique opportunity to directly measure the star mass (Zurlo et al., in prep.).
However, the same data can be exploited to put some constraints on the mass of possible objects around Proxima after calculating the contrast obtained
from these observations.

\section[]{Data and data reduction}

Proxima Cen was observed during six different nights as part of the Guaranteed Time Observations (GTO) program of the SPHERE consortium. The observations are listed in Table~\ref{t:obs}.
All the observations were performed in the IRDIFS mode, with the IFS \citep{Cl08} operating at a spectral resolution R=50 in the wavelength
range between 0.95 and 1.35 \mic with a field of view (FOV) of 1.7\as$\times$1.7\as corresponding to a maximum projected separation from the star of$\sim$1\,au and IRDIS \citep{2008SPIE.7014E..3LD} operating in the H band with the H23 filter pair
\citep[wavelength H2=1.587 \mic; wavelength H3=1.667 \mic;][]{vig10} with a circular FOV with a radius of $\sim$5\as corresponding to a maximum
projected separation of $\sim$7\,au.

\begin{table*}
 \centering
 \begin{minipage}{140mm}
   \caption{SPHERE observations of Proxima Cen. DIT represents the expopsure time for each exposure expressed in seconds, nDIT
     represents the number of frames for each datacube of the dataset. \label{t:obs}}
  \begin{tabular}{c c c c c c c}
  \hline
    Date & Obs. mode & Coronagraph &  nDIT;DIT(s) IRDIS  & nDIT;DIT(s) IFS & Rot.Ang. ($^{\circ}$) & Seeing (\as) \\
         \hline
         2015-03-30  & IRDIFS & N\_ALC\_YJH\_S  &  3$\times$12;16  &  3$\times$12;16 &   3.12  &  0.93  \\
         2016-01-18  & IRDIFS & N\_ALC\_YJH\_S  &  7$\times$40;16  &  7$\times$20;32 &  25.74  &  2.20  \\
         2016-02-17  & IRDIFS & N\_ALC\_YJH\_S  &  11$\times$10;16 &  11$\times$5;32 &  13.52  &  1.86  \\
         2016-02-29  & IRDIFS & N\_ALC\_YJH\_S  &  7$\times$30;16  &  7$\times$15;32 &  22.56  &  0.78  \\
         2016-03-27  & IRDIFS & N\_ALC\_YJH\_S  &  5$\times$40;16  &  5$\times$25;32 &  25.69  &  2.08  \\
         2016-04-15  & IRDIFS & N\_ALC\_YJH\_S  &  6$\times$40;16  &  6$\times$20;32 &  28.72  &  0.62  \\
\hline
\end{tabular}
\end{minipage}
\end{table*}

For both IFS and IRDIS the data reduction was partly performed using the pipeline of the SPHERE data center hosted at OSUG/IPAG in Grenoble.
IFS data reduction was performed using the procedure described by \citet{zurlo2014} and by \citet{mesa2015} to create calibrated datacubes composed
of 39 frames at different wavelengths on which we applied the principal components analysis \citep[PCA; e.g.][]{2012ApJ...755L..28S,2015A&C....10..107A} to reduce the
speckle noise. The self-subtraction was appropriately taken into account by injecting in the data fake planets at different separations.
An alternative data reduction was
performed using the approach described in \citet{2015MNRAS.454..129V} leading to consistent results. 
IRDIS data were reduced following the procedure described by \citet{2016A&A...587A..57Z} and applying the PCA algorithm for the reduction of the speckle noise. An alternative reduction was performed following the procedure by Gomez Gonzalez et al. 2016 (submitted) leading to a comparable
contrast. For all the dataset the contrast was calculated following the procedure described by \citet{mesa2015} corrected taking
into account the small sample statistics as devised in \citet{2014ApJ...792...97M}.

\section{Results}

\begin{figure*}
\centering
\includegraphics[width=0.7\textwidth]{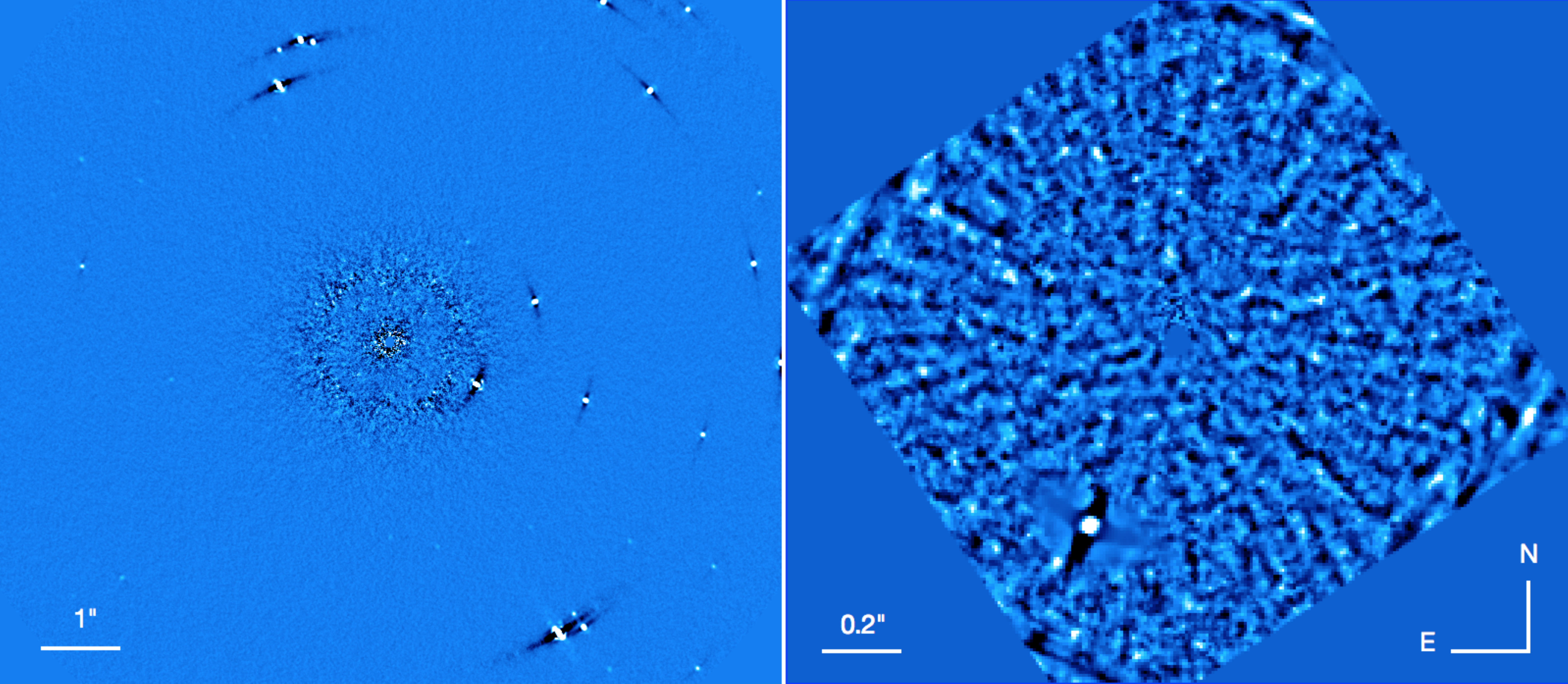}
\caption{Final images obtained for IRDIS (left) and for IFS (right). The IRDIS image is from 2016-04-15 observation while the
IFS image is from 2016-02-29 observation to be able to show the background star that was no more into the IFS FOV in April.}
\label{f:finalimage}
\end{figure*}

Given the very low galactic latitude of Proxima,
several sources were visible in the IRDIS FOV. One example of the reduced images is shown in the left panel of
Figure~\ref{f:finalimage}. Background stars move rapidly in these images due to the large parallax and proper motion of Proxima, so that they can be very easily identified.
One single background source (that is the star undergoing the microlensing event) was visible in the IFS FOV for three observing nights and it is shown
on the right panel of Figure~\ref{f:finalimage}. However, all the detected sources are background stars not bound with Proxima, so that no reliable companion
candidate is detected in the SPHERE images. \par
\begin{figure}
\centering
\includegraphics[width=\columnwidth]{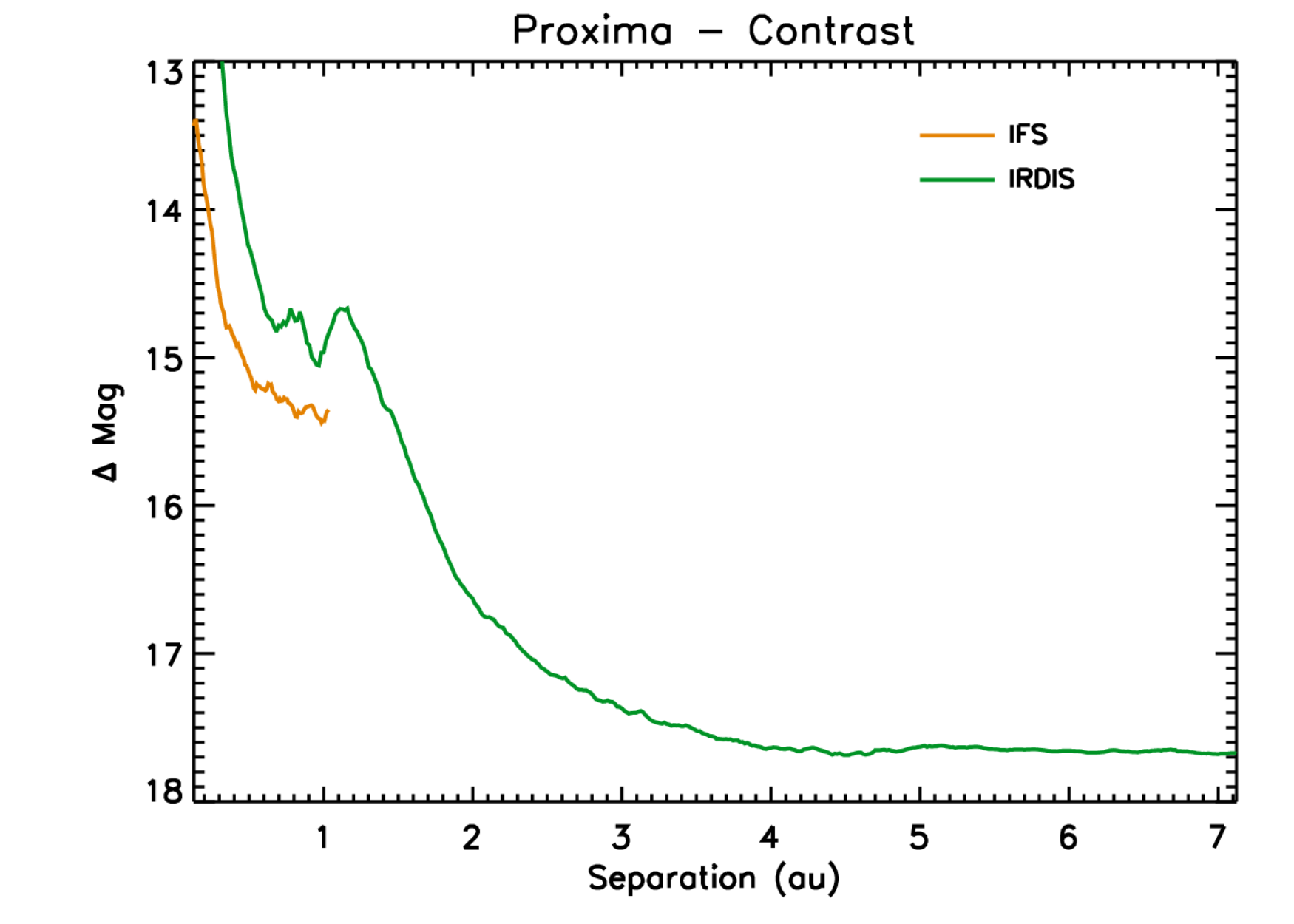}
\caption{Magnitude contrast plot obtained for Proxima using SPHERE. The orange line represents the contrast using IFS while the green
line represents the contrast obtained using IRDIS.}
\label{f:contrast}
\end{figure}

Given the quality of the atmospheric conditions with respect to the other nights (see Table~\ref{t:obs}), the data from the night of 2016-04-15 give the best contrast as shown in Table~\ref{t:contrast}.
Exploiting the very good conditions of this night, we were able to obtain a very deep $5\sigma$ contrast.
As listed in Table~\ref{t:contrast}, the contrast is better than $10^{-6}$ at a separation of 0.4\as using IFS and just above $10^{-6}$ at the same separation
using IRDIS. These values are in good agreement with what expected when SPHERE is observing a very bright target \citep[see e.g.][]{zurlo2014,mesa2015} and is similar to what obtained until
now during the SPHERE observations for targets with similar magnitude \citep[see e.g.][]{2015MNRAS.454..129V}. In Figure~\ref{f:contrast} we display the
contrast in magnitude versus the separation expressed in au for both instruments. We can get a contrast better than 15 magnitudes at projected separations larger than 0.5\,au with IFS while
with IRDIS we obtain a contrast better than 14 magnitudes at the same separation than IFS and we obtain a contrast better than 17 magnitudes at separation larger than of 2.5\,au. \par
\begin{figure}
\centering
\includegraphics[width=\columnwidth]{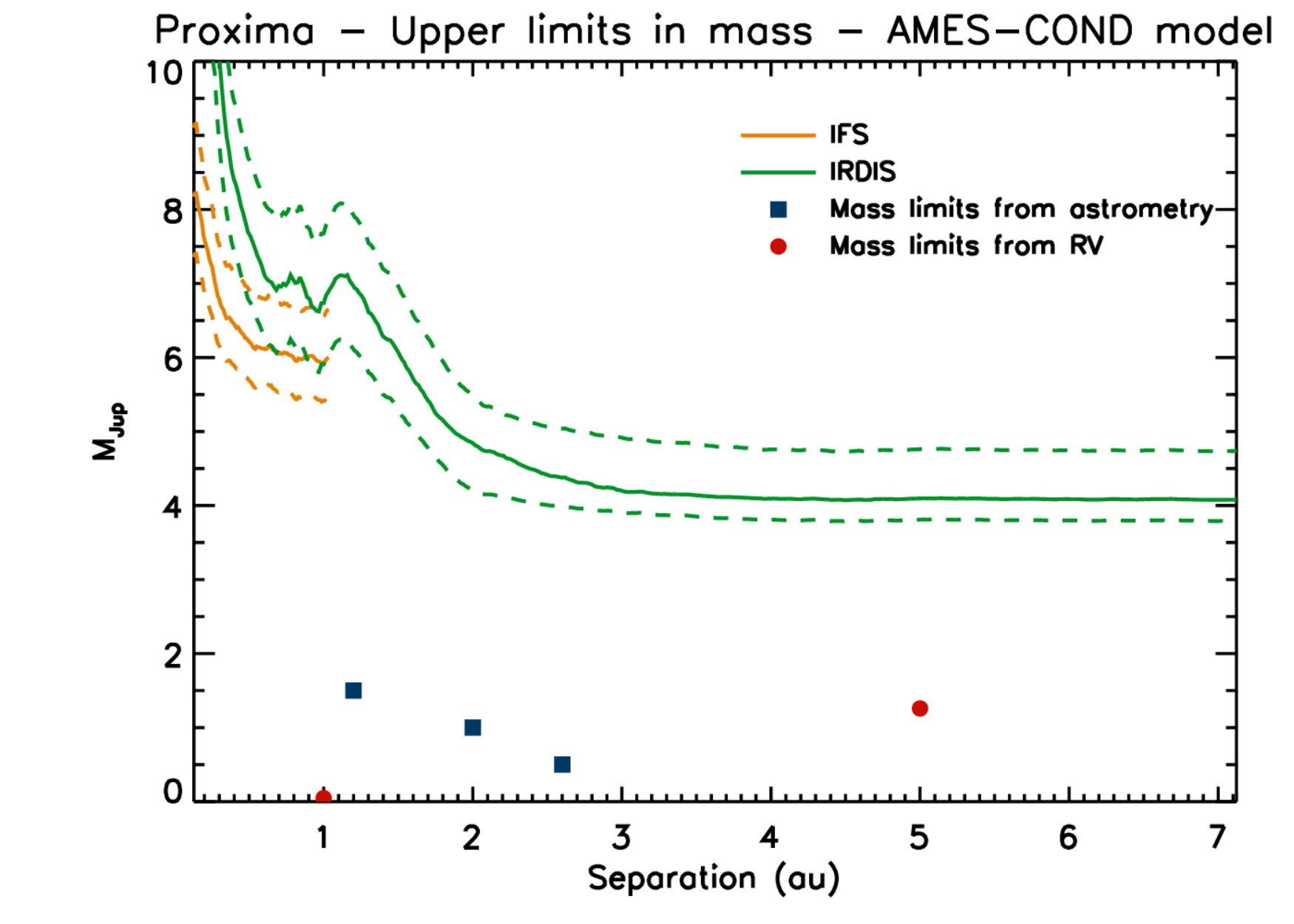}
\caption{Mass limits for planets around Proxima calculated from the SPHERE contrast using the AMES-COND model. The orange lines represent limits from IFS, the green
  ones limits from IRDIS. The dashed lines are drawn to take into account the uncertainties on the stellar age. The limits from astrometry (blue squares)
and the limits from RV (red circles) are also shown as a comparison with our results.}
\label{f:masslimit}
\end{figure}

Using the theorical model AMES-COND \citep{2003IAUS..211..325A}
we were able to set upper limit on the mass of possible objects around Proxima. For this aim we assumed a distance for Proxima of 1.295 pc
\citep{2007A&A...474..653V} and an age of 4.8 $\pm$ 1 Gyr \citep{2002A&A...392L...9T,2016MNRAS.460.1254B}. Moreover, we assumed J and H magnitudes of 5.357 and 4.835 \citep{2003yCat.2246....0C}
respectively for the star.
The upper mass limit plots obtained in this way are displayed in Figure~\ref{f:masslimit} as solid lines. We found a mass limit of $\sim$7.5\,\MJup
at a separation of 0.2\,au and of $\sim$6\,\MJup at separation larger than 0.6\,au with IFS. On the other hand, using IRDIS we were able to get a limit of
8\,\MJup at $\sim$0.4\,au and lower than 5\,\MJup at separation larger than 2\,au. Given the large uncertainties on the age of Proxima, we also calculated the mass
limits considering an age of 3.8 and 5.8 Gyr with the aim to show how the mass limits change according to the stellar age and to set
a more reliable range of mass limits. These results are shown
as dashed lines in Figure~\ref{f:masslimit}. In the same Figure we included, as a comparison with our results, the mass limit obtained by \citet{2008A&A...488.1149E}
using the RV method (shown as red circles) and the limits obtained by \citet{2014AJ....148...91L} using astrometric measurements (blue squares). \par
\begin{figure*}
\centering
\includegraphics[width=0.9\columnwidth]{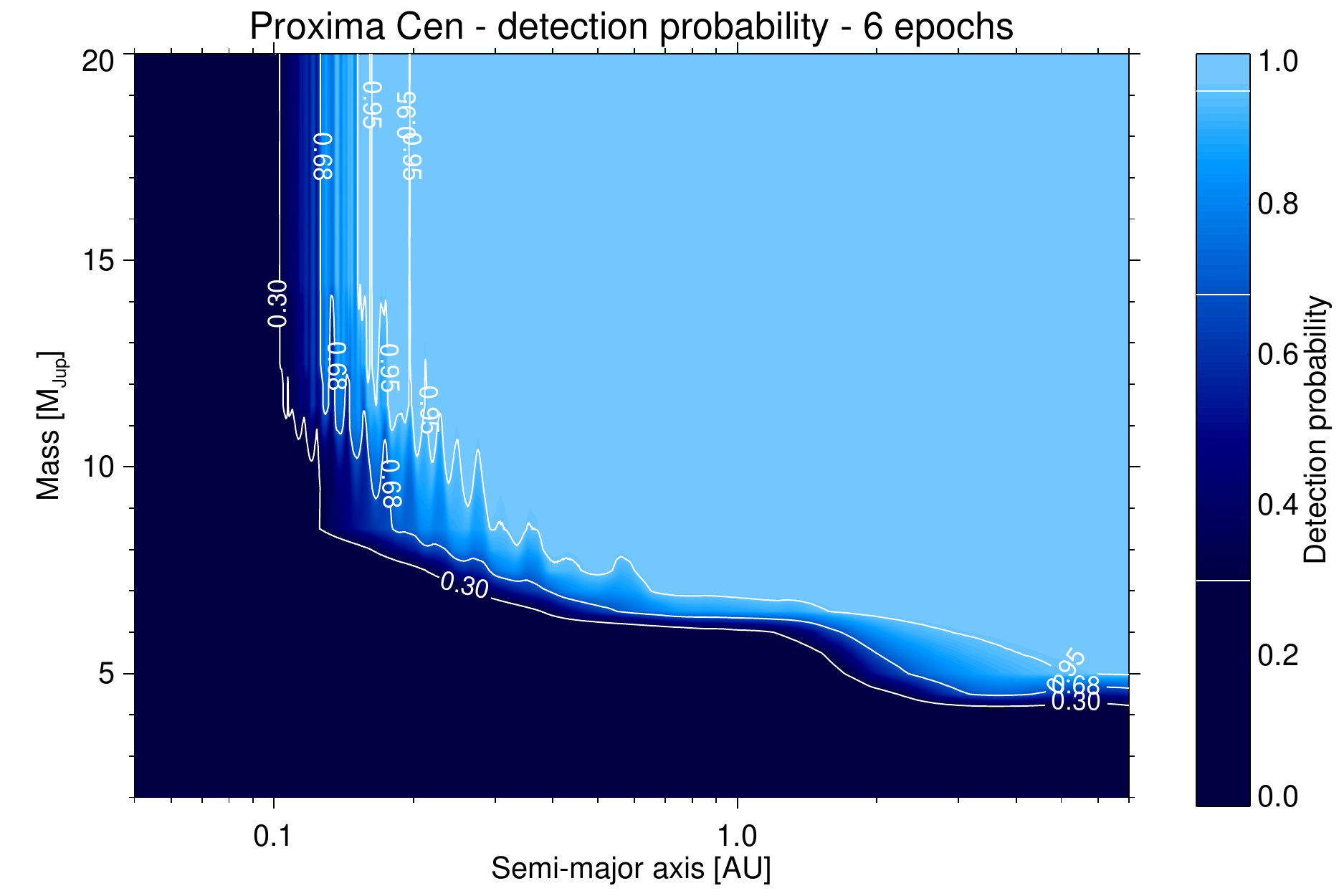}
\includegraphics[width=0.9\columnwidth]{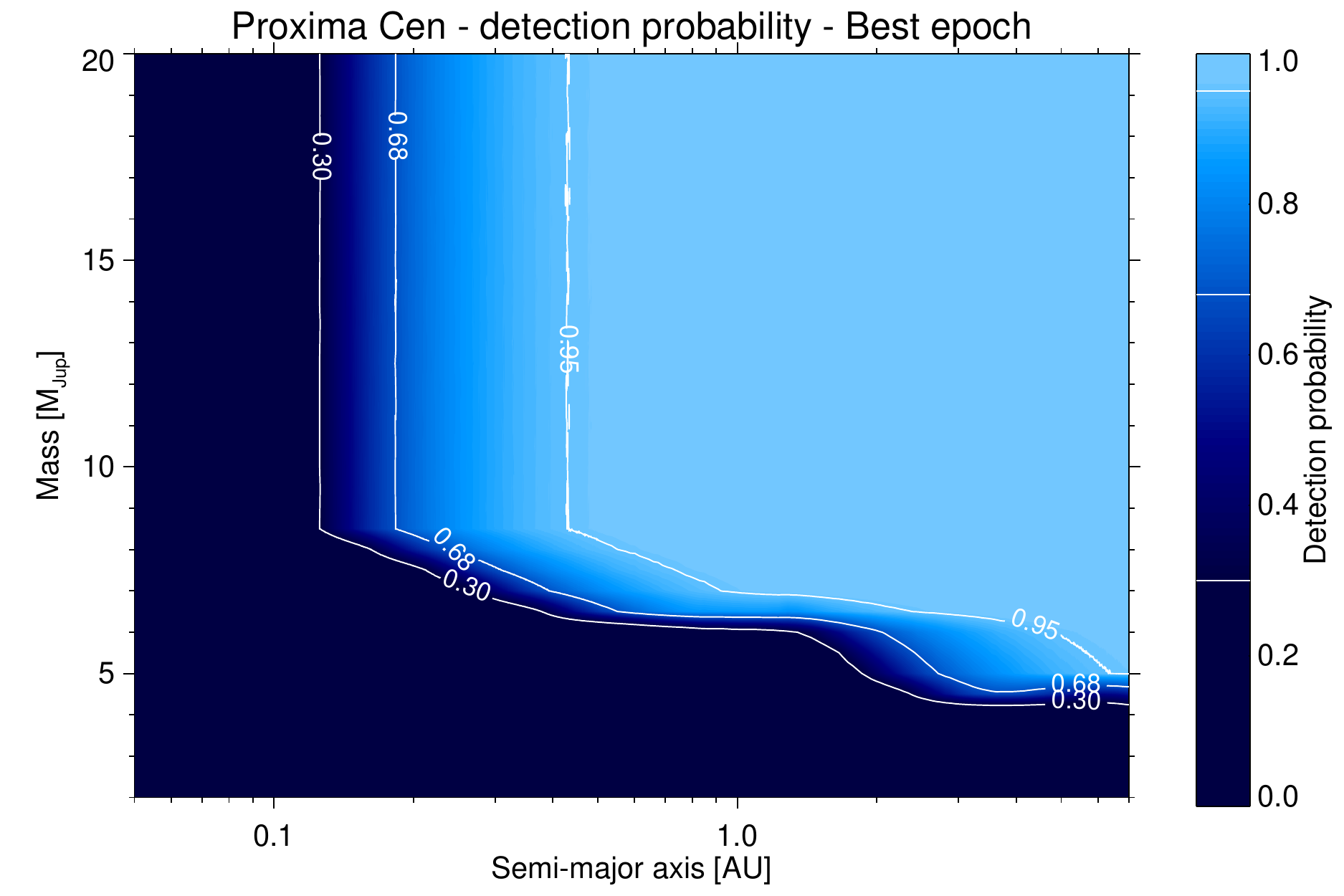}
\caption{{\it Left} Mean probability of at least one detection of a substellar companion around Proxima Cen using the combination
of all the observations as function of the companion mass and the semimajor axis. {\it Right} Same but using just best epoch data.}
\label{f:masslimittot}
\end{figure*}

While the April 2016 data are clearly the best dataset that we obtained, we also combined the data from all the observing
epochs attempting to increase the detection capability. This was performed using the procedure described in \citet{2015MNRAS.454..129V} and based on the MESS program \citep{2012A&A...537A..67B} that is able to
determine the probability of at least one detection in our observing dates calculated on a grid of values for the semi-major axis and for the companion mass. The
results are shown on the left panel of  Figure~\ref{f:masslimittot}. They are in good agreement with the results obtained in Figure~\ref{f:masslimit} but at shorter separations
we are able to obtain a better sensitivity as demonstrated by a comparison with the results of the same procedure performed using only the best epoch data dipslayed on the right
panel of Figure~\ref{f:masslimittot}. This demonstrate, for example, that in this second case the 95\% of probability of detection is cut at 0.4\,au while using all the observations combined we 
arrive at 0.2\,au.  \par 
It is also possible to make an estimation of the limit in radius around Proxima assuming planets shining in reflected light.


However, the contribution to the luminosity of the planet in the regime around Proxima should be mainly dominated by the its intrinsic luminosity
  while the contribution from the reflected light should be less important.
  For this reason the limits obtained through the reflected light are not very meaningful with values ranging  from $\sim$1.5\,\RJup at 0.2\,au to
  $\sim$3\,\RJup at $\sim$1\,au and $\sim$10\RJup at $\sim$7\,au. Values of the radius of the order of $\sim$1\RJup as foreseen from the theorical models are
  then much more probable for substellar objects around Proxima.
%
%
%
%
%
\begin{table}
 \centering
  \caption{SPHERE IFS and IRDIS contrasts at a separation of 0.4\as for the different observing nights.\label{t:contrast}}
  \begin{tabular}{c c c}
  \hline
    Date & IFS Contrast@0.4\as & IRDIS Contrast@0.4\as \\
         \hline
         2015-03-30  &   8.01$\times10^{-6}$  &  8.98$\times10^{-5}$ \\
         2016-01-18  &   5.55$\times10^{-6}$  &  1.03$\times10^{-5}$ \\
         2016-02-17  &   3.82$\times10^{-6}$  &  2.30$\times10^{-5}$ \\
         2016-02-29  &   1.79$\times10^{-6}$  &  5.84$\times10^{-6}$ \\
         2016-03-27  &   3.83$\times10^{-6}$  &  7.22$\times10^{-6}$ \\
         2016-04-15  &   8.58$\times10^{-7}$  &  1.84$\times10^{-6}$ \\
\hline
\end{tabular}
\end{table}
%
%
%

%
%

\section{Discussion and Conclusions}

We presented the results of the analysis of the SPHERE data for Proxima Centauri. While it was not
possible, as expected, to retrieve any signal from the planet recently discovered with the RV technique by \citet{anglada}, we were
able to set constraints on the mass and on the radius of other possible planets around this star. \par
Previous works put constraints on the minimum mass through the RV technique. One example is the
value of $\sim$15 $M_{\oplus}$ at a separation of 1\,au for the minimum mass (M$\sin{i}$)
given by \citet{2008A&A...488.1149E}. Other authors gave similar results. The comparison of these limits with those
obtained by direct imaging allows to exclude face-on orbits for possible sub-stellar objects. 
Extrapolating the results reported by \citet{2009A&A...505..859Z}\footnote{{\bf The time span for these observations was of around 7 years; for comparison the foreseen
orbital period for a planet orbiting at 7\,au is of $\sim$41 years.}} at larger separations, we can conclude that our results
are consistents with those from RV at separation of $\sim$7\,au, that is just at the limit of the IRDIS FOV. \par
Different results were obtained with astrometric measurements. For example, \citet{2014AJ....148...91L} set a mass limit of $\sim$ 1.5\,\MJup
at 1.2\,au, of $\sim$ 1\,\MJup at 2\,au and of $\sim$0.5\,\MJup at 2.6\,au. These constraints are more sensitive to smaller planets than those that
we can obtain with SPHERE. Indeed, we get mass limits of $\sim$7.5\,\MJup at a separation of 0.2\,au and
of the order of 6\,\MJup at separation between 0.6 and 1\,au. However, the wider IRDIS FOV allowed to obtain mass limits at even larger separations
where RV and astrometry are less sensitive. We obtained mass limits better than $\sim$4\,\MJup at separations larger than 2\,au.
It is important to stress that these limits at separations larger then $\sim$5\,au concern a region unconstrained so far.
Moreover, as pointed out by \citet{2011ASPC..448..111D}, model-based substellar mass determinations could be overestimated.
For this reason the mass limits from direct imaging could be even lower than those determined with our measures.\par
We also attempted to obtain a limit for the radius using the reflected light. However, the limits that we obtained are not very stringent probably
because the intrinsic luminosity is more important than reflected light for objects around Proxima. Limits of $\sim$1\RJup foreseen through
the theoretical models are probably more reliable for these substellar objects around Proxima.\par
We obtained these results with a total exposure time of $\sim$1 hour. Given that the residual noise from our observation is mainly dominated by the photon noise
at separations largere than 0.3\as for IFS and at separations larger than 1.5\as for IRDIS, 
we should be able to further improve our contrast with longer exposures taken in sky conditions comparable to those of April 2016 or
better. Under these assumptions, we could be able to improve our contrast as the square root of the exposure time ratio. However, this improvement will not
be comparable with the sensitivity reached by the other methods reported above. For example, a long exposure of 20 hours taken during more than one night will enable us to reach
a contrast of 1.89$\times10^{-7}$ with IFS corresponding to a mass limit of 4.9\,\MJup at a separation of 0.5\as, still far from the limits obtained with RV and astrometry.
To be able to further improve the mass limit obtained with direct imaging we will have to wait for the availability of future instruments both in space (like e.g. {\it James Webb Space Telescope} - JWST)
and from ground using future giant segmented mirror telescope like Giant Magellan Telescope \citep[GMT -][]{2008SPIE.6986E..03J}, the Thirty Meter Telescope \citep[TMT -][]{2008SPIE.7012E..1AN}
and the European Extremely Large Telescope \citep[E-ELT -][]{2007Msngr.127...11G}. \\
Using the online ETC for JWST\footnote{\url{https://devjwstetc.stsci.edu/}} we have calculated the contrast at different separations in the L$^{\prime}$ band for one hour
observation and trasformed it in mass limits using again the AMES-COND models. We synthetized these results in Table~\ref{t:jwst} from which one can see that we can have a
quite good gain especially at larger separation where we can obtain limit similar to those obtained through the RV.  

\begin{table}
 \centering
  \caption{Contrast and mass limit for the L$^{\prime}$ band with JWST.\label{t:jwst}}
  \begin{tabular}{c c c}
  \hline
    Separation (\as) & Magnitude limits & Mass limits (\MJup) \\
         \hline
         0.5  &   20.2  &  5.3 \\
         1.0  &   21.7  &  3.5 \\
         1.5  &   22.6  &  2.5 \\
         2.0  &   23.6  &  1.7 \\
         2.5  &   23.8  &  1.5 \\
\hline
\end{tabular}
\end{table}

\section*{Acknowledgments}
Based on observations made with European Southern Observatory (ESO) telescopes at the Paranal Observatory in Chile, under program IDs 095.D-0309(E), 096.C-0241(G), 096.D-0252(A), 096.C-0241(H), 096.C-0241(E) and 097.C-0865(A).
We are grateful to the SPHERE team and all the people at Paranal for the great effort during SPHERE early-GTO run. D.M., A.Z., R.G., R.U.C., S.D., E.S. acknowledge
support from the ``Progetti Premiali'' funding scheme of MIUR. 
We acknowledge support from the 
French ANR through the GUEPARD project grant ANR10-BLANC0504-01. Q.K. acknowledges support from the EU through ERC grant number 279973.
J.H. is supported by the GIPSE grant ANR-14-CE33-0018. HA acknowledges financial support by FONDECYT, grant 3150643, and support from the Millennium Science Initiative
(Chilean Ministry of Economy) through grant RC130007.
SPHERE was funded by ESO, with
additional contributions from CNRS (France), MPIA (Germany), INAF (Italy),
FINES (Switzerland) and NOVA (Netherlands). SPHERE also received funding
from the European Commission Sixth and Seventh Framework Programmes as
part of the Optical Infrared Coordination Network for Astronomy (OPTICON)
under grant number RII3-Ct-2004-001566 for FP6 (2004-2008), grant number
226604 for FP7 (2009-2012) and grant number 312430 for FP7 (2013-2016).

\bibliographystyle{mn2e}
\bibliography{proxima}


\label{lastpage}


\end{document}